\documentclass[conference]{IEEEtran}

\usepackage{graphicx}
\usepackage{textcomp}
\usepackage{xcolor}
\usepackage{booktabs}
\usepackage{microtype}
\usepackage{algorithm}
\usepackage{algpseudocode}
\usepackage{siunitx}
\usepackage{tikz} 
\usepackage{pgfplots}
\pgfplotsset{compat=newest} 
\usepgfplotslibrary{units} 

\AtBeginDocument{%
  \providecommand\BibTeX{{%
    \normalfont B\kern-0.5em{\scshape i\kern-0.25em b}\kern-0.8em\TeX}}}

\begin{document}
\title{Silently Disabling ECUs and Enabling Blind Attacks on the CAN Bus}

\author{
{\rm Matthew Rogers} \\
University of Oxford \\
matthew.rogers@cs.ox.ac.uk
\and
Kasper Rasmussen\\
University of Oxford\\
kasper.rasmussen@cs.ox.ac.uk}




 \maketitle


\begin{abstract}

The CAN Bus is crucial to the efficiency, and safety of modern vehicle infrastructure. Electronic Control Units~(ECUs) exchange data across a shared bus, dropping messages whenever errors occur. If an ECU generates enough errors, their transmitter is put in a bus-off state, turning it off. Previous work abuses this process to disable ECUs, but is trivial to detect through the multiple errors transmitted over the bus. We propose a novel attack, undetectable by prior intrusion detection systems, which disables ECUs within a single message without generating any errors on the bus. Performing this attack requires the ability to flip bits on the bus, but not with any level of sophistication. We show that an attacker who can only flip bits 40\% of the time can execute our stealthy attack 100\% of the time. But this attack, and all prior CAN attacks, rely on the ability to read the bus. We propose a new technique which synchronizes the bus, such that even a blind attacker, incapable of reading the bus, can know when to transmit. Taking a limited attacker's chance of success from the percentage of dead bus time, to 100\%. Finally, we propose a small modification to the CAN error process to ensure an ECU cannot fail without being detected, no matter how advanced the attacker is. Taken together we advance the state of the art for CAN attacks and blind attackers, while proposing a detection system against stealthy attacks, and the larger problem of CAN's abusable error frames.

\end{abstract}

\section{Introduction}


The CAN protocol was developed in the 1980s~\cite{can20}, and operates most modern vehicles: from consumer automobiles, to trucking~\cite{truckHacking}, to military systems~\cite{military_CAN}. Computers, known as Electronic Control Units (ECUs), are connected to the physical components of the vehicle, and share information across a common serial data bus. For over a decade academia has published work on how insecure the CAN bus is~\cite{ieee_automotive_security},\cite{usenix_automotive_attack}. It has no authentication on any messages, allowing an attacker capable of transmitting on the bus to have control over the vehicle. In response to this lack of security, researchers published security systems, and intrusion detection systems~\cite{Cho:2016:FEC:3241094.3241165}, \cite{choi_voltage_IDS}, \cite{Pawelec2017}, \cite{physical_invariance}, \cite{Sagong2019}, \cite{cryptoJ1939}, \cite{libraCAN}. In this paper we propose a stealthy attack which bypasses these detection systems by producing no visible impact on the bus; no extra messages, no error frames, no changes in timing. This new attack abuses the CAN error process in combination with bit flip attacks to silently force an ECU into a state known as `bus-off'.

ECUs will enter the bus-off state, which turns off their transmitter, after a number of erroneous messages.  The number of errors necessary to reach this bus-off state vary, with a minimum of 32, but we will expand on how the entire CAN error process works in Section~\ref{sec:background}. Regardless, if an attacker has a way to produce errors, then they can control the bus by interrupting messages, and eventually having victims enter the bus-off state. This allows that attacker to effectively replace the victim on the bus, similar to a supply chain attack. 

We propose a novel attack which allows an attacker to push a victim into the bus-off state without producing any errors on the bus. It does this by flipping multiple bits in a single message, rapidly increasing the error count while overriding their victim's error messages and setting the appropriate bits such that other ECUs register it as a valid CAN message. However, the technology to flip bits is inherently complex and may not be 100\% accurate. We introduce a probabilistic attacker, which has a non-100\% chance of success at flipping any particular bit. Through this imperfect attacker we can understand how reliable an attacker needs to be at flipping bits to take a victim off the bus, and avoid producing visible errors. We found that even with just 38\% accuracy for flipping bits, an attacker could silently take a single ECU off the bus within a single message with over 99\% accuracy. More details on this probabilistic attacker are in Section~\ref{sec:probabilistic_attack}.

This all assumes the attacker is in a position where they can read the bus, wait for the appropriate time, then deliver the attack while modifying it on the fly. But this is not true for all attackers. Limited ECU exploits may only allow an attacker to transmit a fixed payload, as they do not have a  large enough buffer to adapt the attack to the bus. Or an electromagnetic induction attack may remotely flips bits on the line with no way of knowing the impact of those bit flips, or even if the flips succeeded~\cite{ElectroMagneticInduction}. As a result many of these attacks rely on the bus being silent when the attack begins so they do not collide with other messages, and mangle their own attack. We propose a new methodology to synchronize the bus for blind (write only) attackers, such that no matter what state the bus is in, the bus is manipulated into the start of a CAN packet. This changes the chance of success for blind attackers from the percentage of dead bus time, to 100\%. Making blind attacks viable threats against the CAN bus.

Given how effective and stealthy our weak probabilistic attacker is, it is clear we need some way of detecting, or in some way preventing bit flip attacks.  Existing intrusion detection systems provide three options: timing analysis, data analysis, and voltage analysis. Timing analysis is ineffective, as bit flips produce no timing differences between messages~\cite{song_interval},~\cite{Cho:2016:FEC:3241094.3241165}. Data analysis gambles on how much the attacker is changing the data, and is inherently limited when the attacker controls every message the victim transmits~\cite{physical_invariance}, \cite{Wasicek2017}. Voltage analysis is the most likely to succeed, but is historically prone to false positives in varied environments~\cite{choi_voltage_IDS}, \cite{cho2017viden}, \cite{Scission}, \cite{Foruhandeh2019} requiring re-training which can then be exploited~\cite{Bhatia2021}. More importantly, the exact tuning for how sensitive the voltage based intrusion detection system needs to be is a constant cat and mouse game based on evolving attacker capabilities. The ideal security update limits how the attack can be done, such that detection is guaranteed. 

The inherent problem our stealthy bus off attack is exploiting is that interrupting an error results in another error. We propose changing the error process such that each error only counts as one error. Meaning that if an attacker causes an error on the bus, that error will continue forever until it is allowed to complete. Our small change has no impact to the safety of the bus, but changes how effective and how stealthy the attacker can be. By having error frames go until completion the attacker now must be 99\% accurate at flipping bits to have even an 80\% chance at completing a single CAN compliant message. And now has to allow error frames to complete to enter the bus-off state.

By forcing the attacker to complete error frames to enter the bus-off state we effectively prevent our stealthy attack; ensuring no ECU can be completely disabled without providing an indicator of compromise on the bus. To account for error prone systems we propose two novel heuristic approaches for differentiating between error frames produced by an attacker, and those produced by malfunctioning systems. As well as a method for detecting an attacker endlessly controlling the bus with bit-flip attacks.


We summarize our contributions as follows:
\begin{itemize}
    \item A novel attack allowing an attacker to silently disable an ECU within one message by flipping bits. As well as a statistical analysis showing they only need 38\% accuracy while flipping bits to have over 99\% accuracy in silently disabling a victim ECU.
    \item A novel methodology which synchronizes the bus such that a blind attacker can guarantee the delivery of their malicious CAN messages regardless of the bus state at the time of the attack.
    \item A proposed change to the CAN protocol's implementation of error frames to remove the possibility for silently disabling ECUs. 

\end{itemize}


\begin{figure*}[tbp]
\center{\includegraphics[width=\textwidth]
{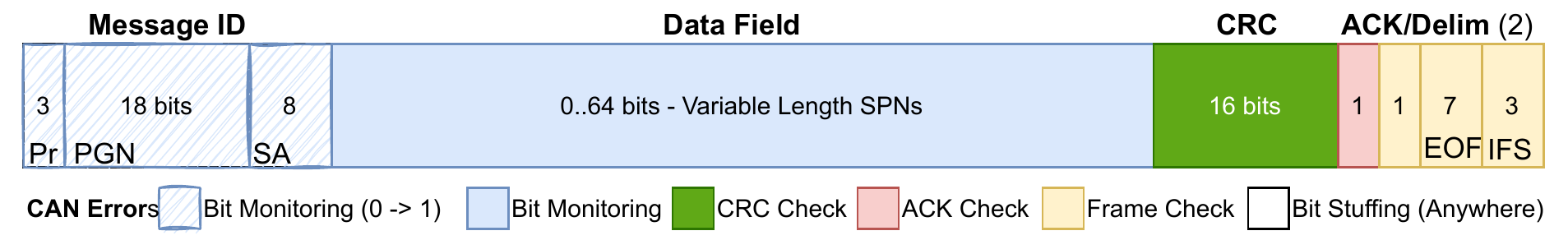}}
\caption{Extended CAN Packet with error sections highlighted where Pr is priority. Bit Stuffing errors can occur anywhere in the packet while other errors are based more on the location. Generally speaking errors occur in response to bit flips, or CRC checks.}
\label{fig:j1939_packet_errors} 
\end{figure*}

\section{Background and Related Work}
\label{sec:background}
In this section we briefly summarize the CAN protocol, with the main focus being on how the CAN error process works. We then go into existing attack and defense work, particularly the bus-off attack which uses a similar technique to our stealthy attack.

\subsection{CAN Background}
CAN is a serial data bus protocol invented in 1986 by Bosch~\cite{can20},  and is used in most modern ground vehicles. The data bus is two 5V differential pair wires, connecting all of the vehicle computers to each others. These vehicle computers, known as Electronic Control Units (ECUs), transmit and receive data on the bus for the purpose of controlling the vehicle in a safer and more efficient way. The packet structure for CAN frames can be seen in Figure~\ref{fig:j1939_packet_errors}. The most notable element for the CAN protocol that separates it from most other message schemes is the arbitration process. ECUs all attempt to transmit at the same time, but back off if they observe a 0 on the bus before they transmit. The short hand is if the ID field is a lower value, then that message has a greater priority, and so wins bus arbitration. For this reason a '1' is referred to as a recessive bit, while a '0' is referred to as a dominant bit. The dominant bit overrides the recessive bit.

\subsection{CAN Error Frames}
At a high level CAN Error Frames are used to cancel out an erroneous message, and inform the rest of the bus that an error occurred. From the bus's perspective this looks like 6 consecutive bits, usually 6 0s, followed by an error delimiter of 8 1s to inform other ECUs the bus is now free,  before finally retransmitting the erroneous message.  The CAN error process works by having each ECU monitor the bus for specific problems, and flag any errors they see. If an ECU registers enough errors, then that ECU turns off its own transmitter to avoid transmitting more errors onto the bus. After receiving enough error-free messages, that transmitter will eventually turn back on, though some implementations require restarting the vehicle to re-enable the ECU. This mechanism is designed for the safety of the vehicle, such that if any ECU frequently talks over other messages, or is unable to properly transmit a signal, it stops inhibiting the rest of the bus. 

Let us examine what CAN considers an error. There are five possible errors~\cite{can_errors_description}: bit monitoring, checksum check, bit stuffing, frame check, and acknowledge check. Figure~\ref{fig:j1939_packet_errors} depicts where these errors occur. \textbf{Bit monitoring} registers an error if the value on the bus is different from what the transmitting ECU believes it sent. \textbf{Checksum check} determines if the 16 bit CRC is valid. \textbf{Bit stuffing} checks that whenever 5 bits of the same value appear, they are `stuffed'. Meaning the opposite bit is transmitted, and the transceiver knows to ignore this bit when assembling the CAN frame. CAN Error Frames use bit stuffing to inform the rest of the bus an error occurred. They do this by transmitting 6 of the same bit in a row without stuffing, raising the bit stuffing error for every ECU so they drop the victim message. \textbf{Frame check} monitors any parts of a CAN frame which are meant to be recessive (1), and transmits an error if they are instead dominant (0). Finally the ~\textbf{Acknowledge check} transmits an error if the ACK field is not set to a dominant bit. Taken together these errors are monitoring tools for when an ECU is no longer able to determine when to transmit, consistently set bits, or calculate a checksum.  

Each ECU tracks the number of errors it sees via internal error counters. Generally speaking each error results in that counter increasing by 8 for the erroring ECU, while each error-free message decrements the counter by 1. If the counter reaches 127 then error frames transmit recessive bits instead of dominant. This is called the passive error state. Passive errors make it easy for a transmitting ECU to talk over the error frames and complete a message under the assumption the error prone device is in the wrong. Interrupting an error frame increments this counter by 8 again. If the error counter reaches 255 the ECU enters the bus-off state, turning off its transmitter. While this mechanism may make sense for safety purposes, it is a clear attack vector.

\begin{figure*}[tbp]
\center{\includegraphics[width=\textwidth]
{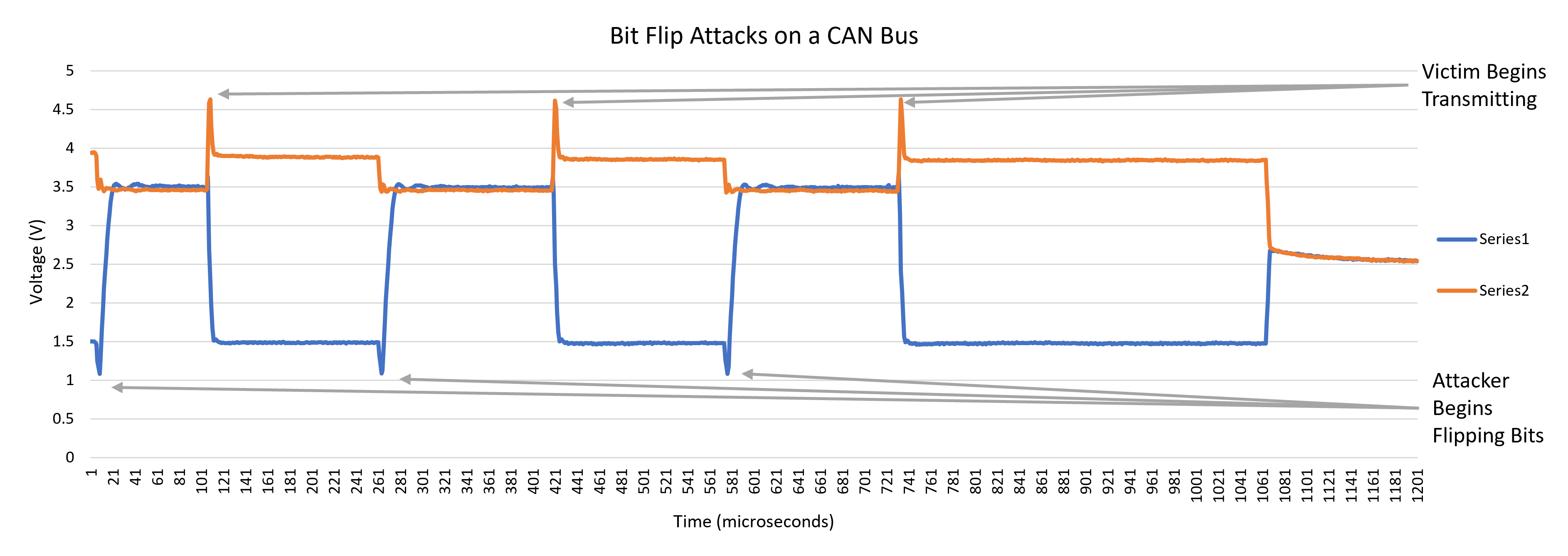}}
\caption{Three extended bit flips attacks on a CAN Bus. The 2nd and 3rd bit flips occur during an error frame, resulting in another error frame. This demonstrates that an attacker can rapidly produce errors by repeatedly interrupting error frames.}
\label{fig:manipulation_attack} 
\end{figure*}

\subsection{Related Work}
There is a great deal of work on hacking the CAN bus~\cite{ieee_automotive_security}, \cite{usenix_automotive_attack},~\cite{truckHacking},~\cite{Sagong2018}~\cite{civic-poc}. Be it remotely~\cite{rail_wifi}~\cite{remoteJeepHack}, or assuming an implanted physical device. Generally it relies on the premise that there is no authentication on the bus, so any message sent has some control of the vehicle. The problem is that then the attacker is conflicting with the existing devices on the bus, and talking louder than them. 

As the field of CAN intrusion detection has matured over the last decade this flood of new attacker messages presents a problem. Several pieces of existing work are designed to detect new messages on the bus, usually through timing characteristics~\cite{song_interval}~\cite{Cho:2016:FEC:3241094.3241165}. Specifically timing intervals between message types, relying on the legitimate ECU transmitting periodically. The way around this is to prevent the legitimate device from speaking, that way the attacker always looks correct. 

There are a few ways to prevent a legitimate device from speaking. Uploading malicious firmware to the victim device is effective for pivoting, but is simple to detect with a robust maintenance process. Industrial CAN standards allow for single network messages to disable a bus, but the attacker must use a very particular message, with a data field never seen in normal traffic, making it easy to detect~\cite{DoS-j1939}.  In terms of silencing an existing device, the best way to do it in CAN is through a `Bus-Off Attack'.

The Bus-Off attack, introduced by Cho and Shin~\cite{off-bus-attack}, synchronizes with a victim message during the arbitration ID, then transmits a dominant bit during the message to produce a bit monitoring error. By producing this bit monitoring error continuously, the bus enters the bus-off state, where the transmitter shuts down. This relies on the assumption that most messages are regularly transmitted, so the attacker can predict when a message will occur, and begin transmitting just before they do so both devices simultaneously win the arbitration process. Significantly, this attack allows error frames to complete. Cho and Shin propose a detection system for Bus-Off attacks based around error frames, but rightfully points out that an attacker could slowly execute this attack, making individual error frames a poor thing to alert on. Other papers expand on the bus-off attack to make it more reliable.~\cite{weeping_can}, but dont fundamentally change the attack. Takada proposes counter attacking a bus-off attacker with a bus-off attack~\cite{counter_bus_off_attack}. However, many CAN controller drivers provide an interrupt to reset the bus-off state, so this only works on limited attackers. In this paper we propose a detection mechanism for the bus-off attack which alerts on exclusively the bus-off state, while also handling a more advanced bus-off attack, capable of flipping any bit on the bus.

Another possible detection system is voltage based fingerprinting. While this technique is good at detecting new devices transmitting on the bus, or attackers transmitting messages they should not be send sending, it does require frequent retraining to account for environmental factors~\cite{choi_voltage_IDS}, \cite{cho2017viden}, \cite{Scission}, \cite{Foruhandeh2019}, \cite{Bhatia2021}. This retraining period creates an easy attack window for our attacker, who can silence an ECU within the span of a single message (within ~300-600 microseconds). Consequently, we believe a mechanism for preventing an ECU from failing silently is necessary, as opposed to a conventional detection system. That said, voltage based fingerprinting solutions would work well in concert with our proposed solution. 
 
\section{System and Adversary Model}
This section outlines how our system is configured, and the adversary's goals and capabilities. Including how we collect traffic, and how our CAN Bus is configured.
\subsection{System Model}

Our system model is a single CAN bus, with any number of ECUs connected to it. We assume all connections to the bus are unbroken, such that an attacker is not man-in-the-middling any traffic. All installed ECUs are CAN protocol compliant. These ECUs encompass features like: powertrain, engine control module, brakes, and transmission, but are not limited to this list. We connect a single device to the bus, acting as a dedicated intrusion detection system (IDS). 

For the purposes of this paper we assume our system is running a basic whitelist IDS, which alerts if any new arbitration IDs appear on the bus. The whitelist contains all arbitration IDs the ECUs on the bus are programmed to transmit. This simple whitelist allows us to detect bit flip attackers attempting to cheat the arbitration process,  or transmit never before seen messages.


\subsection{Adversary Model}
We define an a new attacker called a stealthy attacker. The stealthy attacker is capable of reading and writing to the bus, with a goal of disabling and spoofing ECU(s) without being detected. An attacker unable to meet its goal is deemed unsuccessful.

 We assume flipping a 0 to 1 bit flip has a fixed chance of success from 0-100\%, and a 1 to 0 bit flip has a 100\% chance of success. The justification for this is based on the CAN protocol's defining of 1 as the recessive state, and 0 as the dominant. Such that any ECU transmitting a 0 while another transmits a 1 results in a 0. We expand on the idea of attackers with a not-100\% chance of success in Section~\ref{sec:probabilistic_attack}.

We make no assumptions about attack vector. It could be remote, an implant, or a supply chain attack as long as it can perform bit flip attacks. The only restriction is that the attacker's access vector is not the ECU they are attempting to corrupt or take offline. That said, the nature of the attack vector often limits what the attacker can do. Remote attackers especially can have limited payloads which immediately transmit onto the bus. This leads to scenarios where the attacker's payload collides with other messages on the bus, and becomes nullified by a CAN error frame. Our solution lies in Section~\ref{sec:blind_sync} where we propose a new methodology to synchronize the bus. Such that even a blind attacker, unable to read the bus, can execute an attack and transmit a message on the CAN bus 100\% of the time.


\section{Stealthy Bus-Off Attack}
\label{sec:novel_attacks}

This paper introduces a faster, stealthier bus-off attack for our stealthy attacker. In brief, the original bus-off attack produces errors on the bus by performing a single 1 to 0 flip. Each attack increments an error counter by 8, and by doing it 32 times over 32 messages the attacker puts the victim into the bus-off state. The attacker can go slower to make detection more difficult, but the gist is they get the error counter to 255. Now an attacker can spoof that victim without concerns over conflicting messages between themselves, and the victim. 

Our stealthier version of the attack has the same goal of reaching the bus-off state, but can perform 0 to 1 bit flips, and is unwilling to produce errors on the bus. We also assume the stealthy attacker can arbitrarily flip bits, rather than having to win the arbitration process. The basis of this attack is that a CAN error frame, when interrupted, will produce another error frame, incrementing its error counter by 8 each time it is interrupted. Figure~\ref{fig:manipulation_attack} demonstrates this effect by performing 3 extended bits flips. The first produces an error, which we allow to transmit for a few bits before interrupting with another bit flip attack, and then again. Eventually the error frame transmits 6 bits in a row, and is completed. 

It follows that if an attacker can continuously interrupt error frames, then they can reach the bus-off state within a single message by flipping 32 bits. And once the victim is in the bus-off state, it will stop transmitting error frames, regardless of if any of them ever completed. Now our attacker has turned off an ECU without any error frames transmitted on the bus, and no more error frames being sent. The only remaining catch is ensuring no receive errors are produced, both to avoid detection by our receiving IDS, and ensure the rest of the bus processes their malicious message. This means computing the CRC for their manipulating message, and ensuring the message ends with 11 recessive bits. Both of these are trivial once the bus is in a bus-off state, as 1 to 0 flips are guaranteed in our adversary model. Once the message is done transmitting the bus continues operating as normal, but now with the victim ECU offline. The attacker is free to spoof the victim, taking the victim's place as its receiver remains online, keeping the vehicle functioning while the attacker transmits malicious payloads to other ECUs. All of this happens within a single message, approximately 600 microseconds, with no indicators of compromise communicated on the bus. 

Taking a victim offline within a single message, with no errors, ensures no CAN error based detection systems can detect this attack. Detection systems relying on the timing between messages fail, as the attacker can spoof the timing properties of their victim, and the attack creates no timing variation in the victim message~\cite{Cho:2016:FEC:3241094.3241165}, \cite{song_interval}.  Data based detection systems fail to account for small deviations in data controlled by the attacker (i.e an engine control module is the root of truth for engine data)~\cite{Wasicek2017}, \cite{physical_invariance}. And an attacker can time their attack for the frequent retraining windows of voltage fingerprinting detection systems to avoid detection~\cite{Bhatia2021}.

\begin{figure*}
    \centering
    \includegraphics[width=\linewidth]{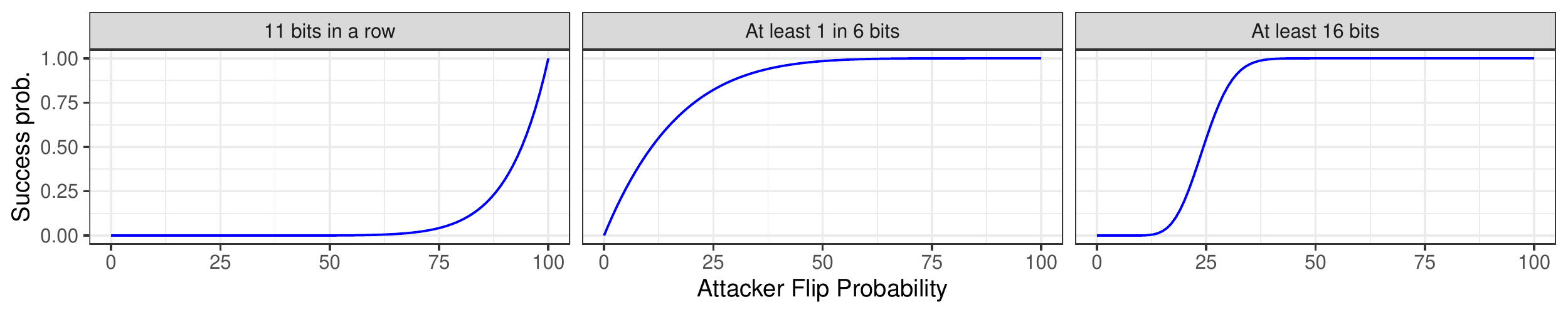}
    \caption{From Left to Right: The probability of a probabilistic attacker to do 11 bit flips in a row. An attacker unable to do this cannot end a CAN packet without generating errors on non-victim ECUs. The probability of a probabilistic attacker to do at least 1 bit flip per 6 bits. An attacker unable to do this cannot reliably prevent errors from being observed on the bus. The probability of a probabilistic attacker to perform at least 16 bit flips in the 64 bit data field. An attacker unable to do this cannot bypass the need to perform 11 bit flips in a row, or guarantee no errors during the arbitration process.}
    \label{fig:probabilistic_attacker}
\end{figure*}

\section{A Probabilistic Attacker}
\label{sec:probabilistic_attack}
Flipping individual bits is inherently a hard problem on a serial data bus, particularly changing the bus from a dominant state to a recessive state like our attacker is doing. In this section we will display how accurate an attacker has to be to successfully take complete control over their victim. Keeping in mind, that being successful requires no indicators of compromise. Meaning not transmitting new arbitration IDs, allowing CAN error frames to complete, or generating errors in receiving ECUs. 

Transmitting new arbitration IDs triggers the simple whitelist run the IDS outlined in our system model. Allowing error frames to complete provides an indicator that something is wrong on the bus, allowing the attacker to eventually be detected. Finally, any errors raised by receivers will also be raised by our IDS connected to the bus, making it immediately obvious an error frame is occurring, regardless of if the attacker interrupts it on the bus. Receiver errors are also problematic because they disable those receiving ECUs. While an ECU is transmitting an error frame, it cannot receive data. This makes sense, errors are coded as 'drop this message' and the erroring ECUs have no reason to process this data. Our stealthy attacker could flip 32 bits to have all receiving ECUs enter the bus-off state, but until they do those ECUs are non-functional. Setting every ECU into the bus-off state is not necessarily a problem if the attacker is prepared to emulate the entire vehicle, but otherwise it is equivalent to a denial of service attack.

Our question is if an attacker with less than a 100\% success rate can avoid either transmitting a new arbitration ID,  completing an error frame, or causing errors in other receiving ECUs. Answering this question reveals how realistic, or unrealistic, our stealthy attack is for attackers with less developed capabilities.


We are concerned with three scenarios that lead to our probabilistic attacker being unsuccessful:
\begin{itemize}
    \item Ability to successfully complete a CAN Packet (11 flips in a row)
    \item Ability to successfully interrupt an error frame (at least 1 flip in 6 bits)
    \item Ability to transmit a known arbitration ID
\end{itemize}


\subsection{Completing a CAN Packet}
The end of a CAN frame requires 11 recessive bits to be sent. This signals every other ECU that another message can now start. If a receiver sees a dominant bit in this period then it will transmit a frame check error frame. The result, is that if an attacker cannot successfully perform 11 dominant to recessive bit flips in a row, then the rest of the bus will stop functioning until an error frame completes or the bus-off state is reached, and our IDS will register a frame check error. The first image of Figure~\ref{fig:probabilistic_attacker} demonstrates the likelihood of flipping 11 bits in a row for different attacker probabilities. We can see that while this attack is possible, if the attacker has less than 94\% accuracy it quickly becomes worse than a coinflip. But, the attacker has another way to more reliably finish CAN frames. If the victim is put into the bus-off state, or error passive state, before the final 11 bits then the victim will be either silent or transmitting a '1'. These are effectively the same thing. The result is twofold: the attacker needs to do nothing to end a CAN frame, and bit flips are easier to execute as the 1 to 0 bit flip is built into the CAN protocol. A standard CAN transceiver can do this flip, so timing concerns aside, we will assume the attacker can always flip a 1 to a 0. 


The only caveat is their ability to perform at least 16 bit flips before the CRC check. The number of bit flips necessary to transmit an updated CRC depends on all prior successful bit flips. However, much like the final 11 bits of the CAN frame, this is trivial is the bus is in a passive error state.
A few system specific bits in the arbitration ID aside, this translates to the ability to flip 16 of 64 bits. The third image of Figure~\ref{fig:probabilistic_attacker} demonstrates that most attackers can do this with a high degree of certainty. The data field may be somewhat nonsensical with lower probability attackers, but after one strange message, they will completely control the victim ECU. 

\subsection{Interrupting an Error Frame}
The ability to interrupt an error frame represents the ability of an attacker to continue transmitting a message without letting an error frame complete, and have the victim message be dropped. This allows an attacker to pretend to be an ECU, without any victim messages being dropped, and eventually putting that ECU in the bus-off state. Fundamentally this translates to an attacker flipping at least one of every 6 bits. The second image of Figure~\ref{fig:probabilistic_attacker} demonstrates that most attackers can interrupt a CAN Error Frame with a high degree of certainty. However this comes with two caveats for our stealthy adversary. 

The first is ensuring receiver errors (CRC, ACK, Stuffing, Frame Check) do not occur, as our IDS can easily detect it. Caveat 2 is based on bus arbitration. We will go into more arbitration issues later in this section, but in this case the main issue is interrupting another ECU attempting to transmit a message. If an attacker flips a 0 to a 1 where another ECU is transmitting a 0, they produce an error frame. Which, again, continues until the attacker lets it complete or the non-victim ECU enters the bus-off state. Both of these caveats are avoidable by having the victim ECU in the bus-off state by the arbitration process, or failing that, in the passive error state. For the purposes of avoiding CRC, ACK, and Bit Stuffing and Frame Check errors the passive error state makes it easy for our attacker to transmit the appropriate bits, as we assume 100\% accuracy for 1 to 0 bit flips. For avoiding arbitration issues, the passive state also makes the arbitration process function normally. Once again, the final image of  Figure~\ref{fig:probabilistic_attacker} demonstrates that reaching the recessive error state is attainable for most attackers by flipping at least 16 bits during the data field. In the case of starting a new message, at least 16 bits in the data field of the previous message.

\begin{figure*}[tbh]
  \center{\includegraphics[width=\linewidth]
    {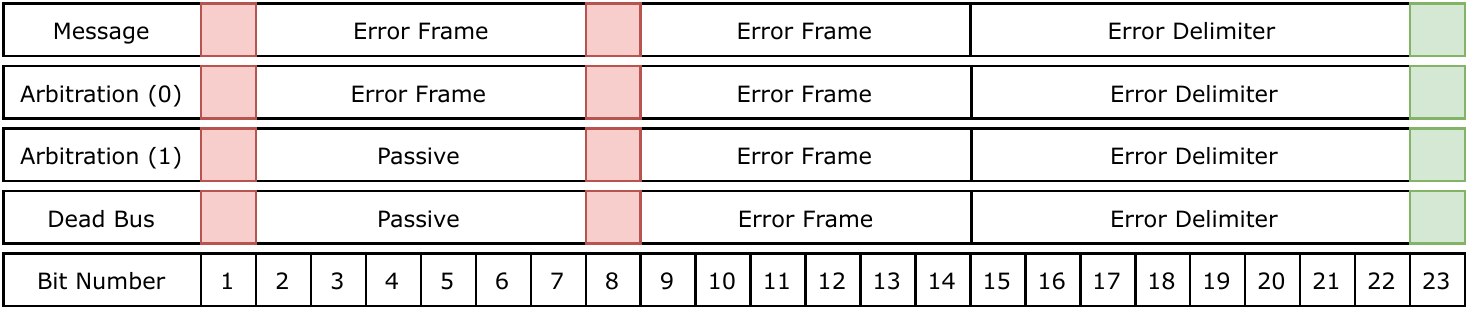}}
  \caption[Blind Can Sync]{Blind Sync Methodology which synchronizes the bus without reading any data, regardless of when the attack is executed. The red boxes at bit 1 and 8 indicate bit flip attacks. The green boxes indicate the point where the bus is synchronized.}
  \label{fig:can_sync}
\end{figure*}

\subsection{Transmitting a Known Arbitration ID}
Finally we must consider the ability to transmit a realistic arbitration ID. If an attacker transmits a message with an ID which has never been seen on the bus before, it is trivial for any on-looker to identify that something is wrong on the bus. It follows that our attacker must transmit an arbitration ID that would be transmitted by the victim ECU. For simplicity we will assume the attacker has a list of all known message IDs. The easiest and safest approach to this problem is to assume a passive bus, all 1s, for the target victim, then perform arbitration normally. Doing otherwise risks accidentally flipping a non-victim ECU's 0 along with an active error frame bit, or potentially failing to flip a particular bit and ending up in a scenario where no valid arbitration IDs exist. We will expand on dominant bit errors during the arbitration process in Section~\ref{sec:bit_flip_attack_detection}.

\subsection{Summary}
From our three scenarios we can conclude that an attacker capable of flipping bits with as low as 38\% accuracy can successfully put a victim into the bus-off state, complete the message, then transmit as that now silent victim over 99\% of the time. All without non-victim ECUs transmitting any error frames, any error frames completing, or any other indicators of compromise on the bus. This shows that this stealthy attack is realistic, even with an adversary with under developed technology that works less than half of the time. The existing error frame process is insufficient for detecting attackers flipping bits on the bus. Thus, we conclude this process needs to be redesigned to handle actual attacks. But before discussing defense mechanisms, we will further prove out the exploitability of the CAN error process by looking at blind attackers and how errors make them more reliable.

\section{Blind Sync Methodology}
\label{sec:blind_sync}
This paper introduces a new methodology of synchronizing the bus for blind, write-only, attackers. The fundamental problem of attacking a serial data bus without reading it first, is that the attacker has no way of knowing what state the bus is in. Lets take the example of a remote attack where the attacker flips arbitrary bits on the bus using electromagnetic interference. Assume we're using a CAN bus with an extended ID field, and 70\% bus utilization.~30\% of the time the attacker transmits a full message, no messages are being transmitted on the bus, and the attack is successfully transmitted. The other 70\% of the time the attacker message is mangled as it collides with another message, and an error occurs. If the attacker continues flipping bits they will likely put that random transceiver into the bus-off state. This is not ideal for actually controlling the victim vehicle, only performing a denial of service. If we want to actually inject an attack, then we need to guarantee that our attack results in transmitting a complete message. To do so we need to make that 30\% chance 100\% by ensuring the bus is silent regardless of what state the bus is in when the attack starts. This means we need a series of bits an attacker can transmit such that regardless of what state the bus is in, we end up at either a silent bus, or the start of a frame.

\subsection{Synchronizing the Bus}
For CAN we can break the possible states into 4 categories: message, arbitration (0), arbitration (1), and dead bus. Figure~\ref{fig:can_sync} depicts bit timings for each of our 4 states, with red representing an attack, and green being the time to start transmitting.
Interrupting a message results in an error frame, after this error frame, an error deliminator occurs for 8 bits, then normal operation occurs again. This is ideal. Interrupt a message once, let the error complete, and then you are confident in the state of the bus until the next error occurs. The trick is lining this error up so it syncs up with the other states. Flipping a 0 bit during the arbitration process causes the same result. However, flipping a 1 bit during arbitration results in the attacker winning arbitration. The attacker cannot effectively use this control of the bus. They have no way of knowing how far into the message they are. Instead, they need to intentionally produce an error to once again try to synchronize the bus. Because they won the arbitration process, there is no other message to interrupt to cause an error frame. Instead they need to stop transmitting for 6 bits. This causes 6 consecutive '1's to appear, resulting in a bit stuffing error. Counting the attack bits, the error frame, and the error delimiter this results in 21 bits. arbitration (0) and message errors only last 15 bits. To sync this up we need to add another bit flip attack. Interrupting any part of a CAN error frame, or the error delimiter results in another error appearing. By adding an attack on bit 8, right as the bit stuffing error frame starts, and just after the bit monitoring error frame ends, we extend the length of the bit stuffing error frame by 1 bit, and create a new error frame for the message and arbitration (0) case.

Now the message, arbitration(0) and arbitration(1) cases are synchronized. This leaves the dead bus case. The attacker cannot simply transmit a message during the dead bus time, as that only works 30\% of the time. Instead they need to intentionally produce errors to sync with the rest of the bus. This case works the same as the arbitration (1) case, except the initial bit flip starts a CAN frame. Considering the bus was silent, no other devices are competing to speak. Thus, the attacker can copy the same technique as arbitration (1): stay silent for 6 bits, and raise a bit stuffing error. These techniques have the same timing property. Now, by flipping a bit, waiting 6 bits, then flipping the next bit, the entire CAN bus is synchronized  until the next error frame.  The only evidence of this attack is a single error frame completing, possibly two depending on the level of information provided by the CAN controller. The blind sync methodology can be used for any attack where the attacker cannot read the bus, and cannot queue messages. 

\subsection{Transmitting An Attack}
Once the bus is synchronized, the blind attacker can transmit any message they wish, with the caveat that they are likely to produce errors unless they win the arbitration process by setting the first few bits to 0. This will create new arbitration IDs on the bus, making the blind attacker easy to detect. But only so far as knowing an attack occurred. traditional CAN logging, and even the CAN logging done in this paper, has no mechanisms for identifying how this attack occurred. The other option is picking a high priority but low frequency message, such that the attack is likely to win the arbitration process, but will not accidentally share that win with the legitimate ECU. Regardless, our blind sync methodology ensures a blind attacker can have their desired impact on the vehicle 100\% of the time, regardless of what is happening when the attack is launched. A vast improvement over the blind attacker's previous 30\% chance of success.

\begin{figure*}[tbh]
  \center{\includegraphics[width=0.9\linewidth, page=2]
    {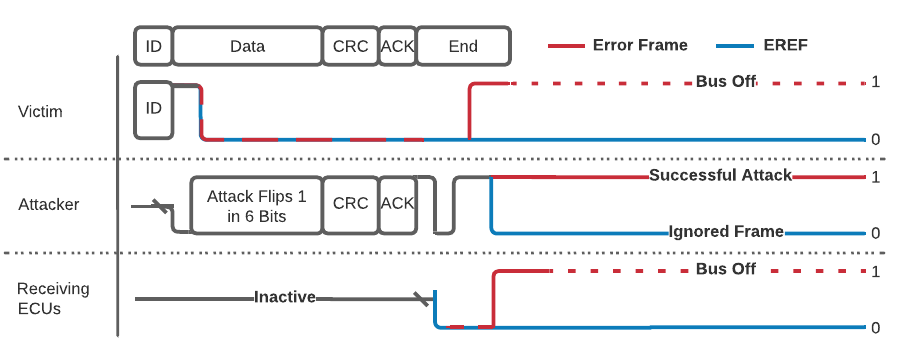}}
  \caption[Cascading Errors]{Demonstrates the effect of an attacker inducing errors onto the bus where red lines indicate normal error frames, and the blue lines indicate error resistant error frames. This shows that an error during the end of a CAN frame results in all receiving ECUs generating errors. The normal error process enters the bus-off state, while EREFs continue transmitting dominant error frames. With receiving ECUs in the bus-off state, the bus will process the attacker payload, while EREFs cause it to be ignored}
  \label{fig:cascading_errors}
\end{figure*}

\section{Bit Flip Attack Detection}
\label{sec:bit_flip_attack_detection}
From our probabilistic attacker section it is clear that most attackers capable of flipping bits on the bus can silently turn an ECU off, without any errors appearing on the bus. The attacker can then seamlessly pretend to be that device, with on-lookers none the wiser. The critical detail here is that the attack is only easy for the majority of attackers because they can put the victim ECU into a passive error state, making it trivial to finish CAN messages, perform a normal arbitration process, and interrupt CAN errors. If we can prevent this passive state then only highly accurate attackers can properly end messages. Based on these conclusions we require a change to the CAN error process to prevent our stealthy attacker from being able to reliably control their victim. One that ensures the detection of any bit flip attacks and prevents attacks from occurring within a single message, even for attackers with 100\% accuracy for flipping bits.



\begin{algorithm}[bt]
\caption{Error Resistant Error Frame Algorithm}\label{alg:cap}
\begin{algorithmic}[1]
\State $lastError \gets 0$
\Procedure{OnError}{}
\If{$lastError > 0$}
\State $pass$
\ElsIf{$lastError <= 0$}
\State $TEC \gets TEC + 8$
\EndIf
\State $lastError \gets 6$
\EndProcedure

\Procedure{OnBit}{$bit$}
\If{$ErrorInBit(bit)$}
\State $OnError()$
\Else{}
\State $lastError \gets lastError - 1$
\EndIf
\EndProcedure

\end{algorithmic}
\end{algorithm}

\subsection{Error Resistant Error Frames}

We propose modifying the error process such that if a CAN error frame is interrupted, it will not increase its error counter until the error frame is completed, and only for a singular error.  We will refer to these as Error Resistant Error Frames (EREFs). We demonstrate an algorithm for EREFs in Algorithm~\ref{alg:cap}, where the `OnBit' procedure is run whenever a bit is received. Our modifications to normal error and bit processing ensures that transmission error counts (TECs) are only increased if the observed error isn't within the transmission period of another observed error.

EREFS drastically change the 3 scenarios outlined in Section~\ref{sec:probabilistic_attack}.  For our first two scenarios, it means the easy way out of having the victim stop transmitting 0s midway through the message is gone. Instead, in our `completing a CAN frame' scenario the attacker must be able to flip 11 bits in a row, or our IDS will detect the frame check error, and no message will complete. Even if they can drive the entire bus, the transceivers would not actually receive data because every receiving ECU started transmitting errors. We demonstrate this cascade of errors in Figure~\ref{fig:cascading_errors}, where EREFs cause all receiving ECUs to continue transmitting dominant error frames, and normal error frames cause receiving ECUs to enter the bus-off state. These continuing dominant error frames from all receiving ECUs mean that EREFs make bit flip attacks unreliable for any attacker that cannot perform 11 bit flips in a row with near 100\% certainty. Based on Figure~\ref{fig:probabilistic_attacker}, we can see that EREFs make ending a CAN Frame without errors unlikely for most attackers, with even a 99\% attack chance only having an 80\% success rate over 11 bits. The story is similar for the `interrupting CAN error frames' scenario. A successful attacker would be assembling a reasonable message, with a valid CRC, while also interrupting 1 of every 6 bits and adjusting on the fly for attacks that do not work. This is easier than ending a CAN Frame, but again, prone to errors. Especially once the attacker reaches the CRC and needs a very specific set of 16 bits, and failure means either detection from an error frame completing, or our IDS seeing a receiver error.

EREFs place particularly strong restrictions on the attacker while trying to pick a valid arbitration ID. We can no longer assume the bus is in error-passive or bus-off mode, so the attacker is actively flipping a '0' to a '1' during the arbitration ID process, even when they aren't attempting to transmit a message from their chosen victim ECU. If at any point another ECU attempts to transmit a 0 and the attacker performs a bit flip, then that ECU will begin to transmit a bit monitoring error frame as its 0 is flipped to a 1. This error ensures that ECU will stop receiving data until the error frame completes. Because the ECU cannot complete an error frame without providing an indicator of compromise, they have no way of re-enabling an ECU transmitting errors. But the attacker must perform bit flips during the arbitration, not only to prevent the CAN error from completing, but because otherwise it will transmit previously unseen arbitration IDs, providing an easy indicator of compromise. However, differentiating between a 0 only transmitted by an error frame, and any number of ECUs transmitting a 0 on top of that error frame is infeasible. Meaning, the attacker must predict the exact order of messages going over the bus or risk other transmitting ECUs generating errors, or an unknown arbitration ID. This is true for even an 100\% attacker, as their reliability does not affect the cascading error problem. The difficulty in starting a message as a result of EREFs demonstrates that EREFs make bit flip attacks infeasible for our stealthy attacker.

\subsection{Detecting Driving the Bus}
\label{sec:driving_the_bus}
An attacker avoiding detection may end up in a scenario where they are continuously driving the bus, as they are unwilling to let error frames complete. Now, many of the ECUs on the bus may not be functioning, as they are transmitting error frames, but the attacker can still transmit messages on the bus as if ECUs were listening. This is a good way of fooling an IDS. While an observer may easily be able to tell the vehicle is not working, the IDS with only bus data has no way to know. Assuming the attacker can emulate every ECU on the bus.

To get around this our IDS needs a way of knowing if an attacker is driving the bus. We propose having the IDS randomly transmit unused CAN IDs with the first 3 bits being 0.  Because the attacker has no way of distinguishing between the 0 of a CAN error frame, and the 0 of our IDS transmitting, they will inevitably flip a bit in our IDS's arbitration ID to 1. The random transmissions times ensure an attacker cannot predict our security message. This results in our IDS generating an error frame, and our attacker being detected.

For the safety of the rest of the CAN Bus, our IDS cannot transmit too frequently. Otherwise legitimate ECUs may not be able to transmit their data in a timely manner. Given most high priority CAN messages repeat periodically within the 10-100 millisecond range,  we believe once every 1 to 5 seconds is slow enough to detect an attacker quickly without impacting bus performance.

\subsection{Completing Error Frames}
\label{sec:error_frame_detection}
Our adversary model assumes that any error frames are an indicator of compromise, and so the attacker is unwilling to let any complete. It follows that EREFs, which practically guarantee cascading errors for an attacker, make bit flip attacks equivalent to denial of service attacks. However, this does rely on the idea that error frames are inherently rare, which varies from system to system. For example, the original bus-off attack paper has to produce at least 16 error frames, and the authors make the assumption that alerting on any individual error frame would result in false positives, as they are somewhat common in normal traffic~\cite{off-bus-attack}. While we have not observed this in our data, we do acknowledge it as a possibility. 

Any attack that produces error frames is inherently more detectable by an incident response process because it contains some indicator, while our stealthy attacker produces none on a normal CAN bus. In this section we will examine how to detect the original bus-off attack, regardless of how slowly the attack is done, and then consider the ramifications of EREFs on our attacker if they are willing to produce errors.

\begin{figure}[tb]
  \center{\includegraphics[width=0.95\linewidth]
    {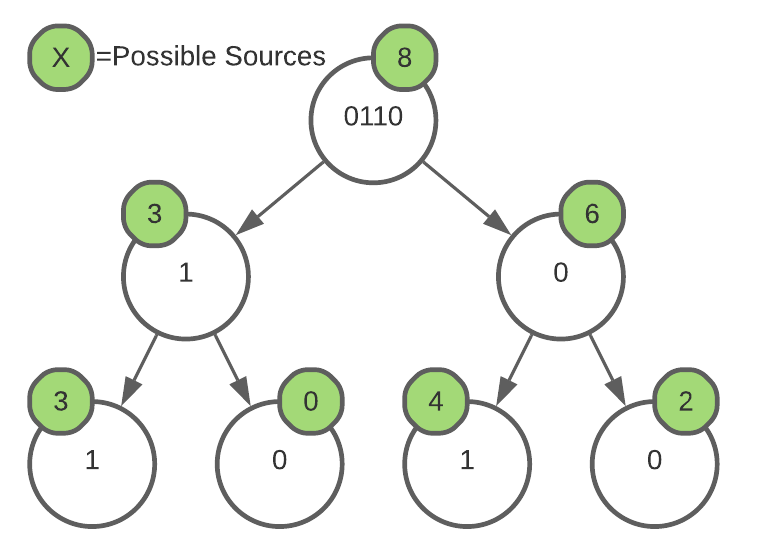}}
  \caption[Identification Tree for CAN]{Tree view indicating how early into a CAN packet one can narrow down the possible source address of the transmitter without any voltage fingerprinting. Starts on the 4th of 29 bits in an extended CAN packet, where bits 21 to 29 indicate the source address. Pulled from a data set with 17 sources.}
  \label{fig:can_tree_figure}
\end{figure}

\subsubsection{Detecting Traditional bus-off Attacks}
For a normal CAN controller, using normal CAN error frames, the only notification of an error occurring on the bus is in a flag at the start of a message. It is not clear who transmitted the error, simply that an error occurred. This is unhelpful, as normal CAN controllers also drop the message with an error in it, so the application layer has no way of conclusively connecting the dots. Without some attribution for errors, it becomes difficult to gauge how many errors any individual LRU is producing, and whether it is close to the bus-off or passive error state. Alerting on errors before these critical point could produce false positives, as an error could be indicative of a one-off bug where a transceiver syncs incorrectly, or a loose connection. But alerting on these critical points is less error prone, as we assume the system functions in normal driving conditions, which it could not do if it regularly moved into the passive error or bus-off states. Either from mangled or illegal messages not being dropped by the passive error ECU, or that ECU not functioning altogether in the bus  off state. 

We use a custom CAN controller which does not drop messages with error frames. This way we can identify what ECU the error is in, and track our own error counters for each ECU. This means we increase the error counter by 8 for each error from an ECU, and decrease it by 1 for every good message from that ECU. Of course, the attacker could initiate a bit monitoring error before the source address is transmitted in the arbitration ID, but we can extrapolate from the bits that were transmitted before the error to determine what ECU was transmitting. Figure~\ref{fig:can_tree_figure} demonstrates this logic by applying it to one of our datasets with over 10 million CAN packets.  There are 17 ECUs initially, by parsing the priority field (the first 3 bits)  we narrow this list down to 8 ECUs. Choosing another 0 narrows the list to 6, and then another gets us to 2 possible source addresses. We can continue this process to eventually narrow the list to 1 possible source address before we enter the source address field. An attacker can use this technique to reliably silence a single victim ECU as early as possible. A defender can use this technique to identify what device was transmitting, or where an error occurred, without any voltage fingerprinting. As simple a technique as counting ECU errors ourselves is enough to detect the original bus-off attack~\cite{off-bus-attack}, even without EREFs. This is because the traditional attack lacks the ability to produce more than one error at once. 

\subsubsection{Error Detection with Multiple Errors}
\label{sec:obfuscating_error_frames}
Now let us consider this from the perspective of our attacker willing to produce errors in other ECUs, with EREFs enabled. At any point the attacker can complete an error frame, and that error frame could include the errors from multiple ECUs. As they could have artificially driven the bus for any period of time. However, the error counters for each of those ECUs will only increase by the standard amount for one error frame because of EREFs. The difficulty is determining which ECUs those errors were introduced in.  But first let us understand why our attacker might want to complete an error frame. While an ECU is transmitting an error frame it cannot receive data. CAN controllers are simply not built around the idea of multiple ECUs transmitting at the same time outside the arbitration field. Now the attacker has two goals: complete 16 error frames so the victim stops transmitting dominant error frames, and allowing error frames to complete so that the vehicle can function until the attacker gains control of the victim. We presume the attacker knows about our proposed error frame based bus-off attack detection, and so will not produce 16 error frames in a row. Instead they will seed their errors into other ECUs by carrying the error frames from their victim into a synthetic message they produce from a non-victim ECU. This brings us back to the the difficulty of determining where these error frames were introduced. By spreading the ending point of the errors around, it becomes unclear which ECU is approaching a dangerous error state.

We identify 2 possible mechanisms for determining which ECUs were transmitting errors frames and detecting an attack. The first relies on the replay functionality within CAN. When a message is interrupted with an error frame, the transmitting ECU immediately replays that message after the error completes. Thus if we see a message re-appear quickly, its indicative that an error was introduced in the message. The attacker could continue generating errors with that replayed message, but unless their target is a particularly low priority message, that means almost no normal ECU traffic occurs from the introduction of one error to the next. And if they do pick a low priority message, then they must listen through a number of bits on the arbitration ID to ensure they are not interrupting a non-victim ECU. At the exact moment an attacker can determine which message is which, we can do the same using the same logic outlined by Figure~\ref{fig:can_tree_figure}. By adapting existing timing based work which predicts the time interval between arbitration IDs~\cite{song_interval}, we can create bands of time where we would expect to see each message, and then identify common bits at the start of the arbitration ID that would indicate an attacker targeting that message or ECU. We chose interval based work as an attacker has no mechanism for spoofing the time between messages. Other non-specialized hardware detection mechanisms such as clock skew analysis~\cite{Cho:2016:FEC:3241094.3241165}, can be faked by an attacker~\cite{Sagong2018}.

The second is based on aggregate error counts. While we cannot alert on any individual error, we can alert on errors occurring more frequently than normal. We assume that if error frames occur frequently enough that they could trigger false positives, then they are deterministic, and can be expected to regularly occur. Meaning if we take a collection of normal traffic and calculate the average number of errors from each ECU, then we can safely alert on any deviations from this number. We take this approach because error counters decrease for every good message transmitted by the erroring ECU. Meaning a static error count is likely to be incorrect, as the attacker could be artificially running the bus for several messages. This means we have full certainty in the number of errors, but not full certainty in the number of error-free messages.

Either of these approaches is effective against an attacker willing to complete errors on the bus. Particularly because an attacker must complete errors quickly, or have parts of the bus stop functioning. This is because EREFs prevent the error counter from incrementing without the error completing. While limited attacks are possible by introducing one off errors, our proposed detection systems and EREFs ensure it is not possible to turn off an ECU without some indicator on the bus. A full implementation of both of these intrusion detection systems is outside the scope of this paper, but the vast landscape of CAN intrusion detection research they build upon gives us confidence in their efficacy.

\section{Security Analysis}
This security analysis will examine what attacks we are capable of detecting regardless of the attacker having full knowledge of how our detection system works. The goal of the stealthy attacker is to remain undetected while disabling ECU(s). In order to disable an ECU they must increment the transmission error counter to greater than 255. Flipping a bit on the bus increases the transmission error counter by 8, and transmits an Error Resistant Error Frame (EREF). The attacker can allow the error frame to complete. This error is seen by our IDS, and allows them to be detected, with more specific error frame alerting outlined in Section~\ref{sec:error_frame_detection} for error-prone busses. Thus the attacker must interrupt the error frame continuously to avoid detection. Because no error frame is ever completed, this error frame is always transmitting dominant bits. If the attacker generates a CRC, Frame Check, or ACK error then our IDS, along with all receiving ECUs, generate an error, resulting in detection. Regardless of if the attacker interrupts this error. This is not a problem for our 100\% success attacker, but is for all other attackers, with only an 80\% chance for the 99\% attacker to not produce a frame check error. 

Now the attacker has completed a frame by continuously interrupting the EREF. At the start of the next frame the attacker must continue flipping the EREF's dominant bits to produce a known arbitration ID, and prevent the error frame from completing. On flipping a bit in the arbitration ID if another ECU intended to transmit a dominant bit, an error occurs. Thus, cascading the original bit flip error into another ECU. On another ECU transmitting an EREF, that ECU stops receiving data. As an attacker cannot complete an error frame without being detected, this results in ECUs being disabled through every arbitration process unless the attacker perfectly predicts which ECUs are going to transmit on the bus.  The attacker cannot perfectly predict the order of messages on the bus, as shown by existing intrusion detection work failing to do so~\cite{song_interval}, thus ECUs must be disabled as lower priority value messages are transmitted on the bus. Eventually our IDS with transmit a randomized arbitration ID with the first 3 bits set to 0, as outlined in Section~\ref{sec:driving_the_bus}. Our IDS will always win this arbitration process in a non-error state. The attacker can not differentiate the 0 from our IDS from the 0 of an EREF, and will flip it to transmit known messages. This flip produces a bit monitoring error in our IDS, allowing us to detect the stealthy attacker. In summary, EREFs ensure our stealthy attacker has no way of succeeding in its goals, as it is always detected.

\section{Discussion}
This paper demonstrates the effectiveness of using the CAN error frame to hide and enable attacks. The fundamental problem is the inability of ECUs to communicate their own error state on the bus. At least, not in a scenario where the attacker can arbitrarily flip bits on the bus. Any attempts to communicate what error state an ECU is in would be overwritten. This can be solved a couple of ways. Adding authentication to the bus means that an attacker cannot overwrite a bit without knowing some secret key. The secrecy of that key is in question for remote attackers since they effectively land on the device, but stealthy bit flip attacks require custom hardware, making it unlikely they could circumvent authentication. While adding authentication also requires modifying ECUs, it also uses up valuable bus bandwidth. The other option is a secondary channel for communicating errors. Again, it requires modifying ECUs, but also comes with the additional wiring costs. Arguably this approach just moves the attack somewhere else. The simplicity of EREFs is the main appeal over these other two techniques. They are a mechanism for ensuring the protocol works as intended. That dominant CAN error frames actually complete. On a normal bus they do not take up bandwidth, or have any visible impact at all. They only appear when an attacker does, and that is the ideal for security systems.

\section{Conclusion}
In this paper we proposed a novel attack, a methodology for enabling blind attackers, a change to the CAN protocol which prevents attackers from stealthily removing ECUs from the bus, and two proposed heuristic techniques for differentiating legitimate error frames from attacker induced error frames. Altogether this paper provides a thorough examination of an attacker capable of flipping bits, and what they can do on a CAN bus. Even if they are incapable of reliably flipping bits. As the CAN protocol exists currently, an attacker can silently disable any number of target ECUs, even with just a 38\% chance of successfully flipping a bit. Our addition of Error Resistant Error Frames then puts the attacker in a position where they have no choice but to inevitably complete an error frame, or have the bus stop functioning. This allows us to detect the attack, while our CAN error detection mechanisms differentiate the legitimate error frames, from attacker induced error frames. Outside of detection we proposed a novel attack methodology which enables blind attackers to always have an impact on their victim, regardless of what the CAN bus is doing as they initiate their attack. Taking their chance of success from the percentage of dead bus time, to 100\%. Through the totality of this work future researchers can more accurately model against bit flip attacks as they consider the ramifications of more advanced attack vectors for the innumerable number of devices using the CAN protocol.

\bibliographystyle{plain}
\bibliography{main}

\end{document}